 \documentclass[prb,10pt,twocolumn,showpacs,showkeys,preprintnumbers,amsmath,amssymb,superscriptaddress,aps]{revtex4-2}

\usepackage{graphicx}
\usepackage{dcolumn}
\usepackage{bm}
\usepackage{amsthm}
\usepackage{amsmath}
\usepackage{amssymb}
\usepackage{xcolor}
\usepackage{siunitx}

\DeclareSIUnit{\angstrom}{\textup{\AA}}

\begin{document}

\title{Micromagnetics of ferromagnetic/antiferromagnetic nanocomposite materials. \\Part~I:~Towards the mesoscopic approach}

\author{Sergey~Erokhin}\email{s.erokhin@general-numerics-rl.de}
\affiliation{General Numerics Research Lab, Kahlaische Str.\ 4, D-07745 Jena, Germany}
\author{Dmitry~Berkov}
\affiliation{General Numerics Research Lab, Kahlaische Str.\ 4, D-07745 Jena, Germany}
\author{Andreas~Michels}
\affiliation{Department of Physics and Materials Science, University of Luxembourg, 162A~Avenue de la Faiencerie, L-1511 Luxembourg, Grand Duchy of Luxembourg}


\begin{abstract}
In the first of two articles, we present here a novel mesoscopic micromagnetic approach for simulating materials composed of ferromagnetic and antiferromagnetic phases. Starting with the atomistic modeling of quasi one-dimensional systems, we explicitly show how the material parameters for the mesoscopic model of an antiferromagnet can be derived. The comparison between magnetization profiles obtained in atomistic and mesoscopic calculations (using a Heusler alloy as an example) proves the validity of our method. This approach opens up the possibility to recover the details of the magnetization distribution in ferromagnetic/antiferromagnetic materials with the resolution of a few nanometers covering length scales up to several hundreds of nanometers.
\end{abstract}

\maketitle

\section{Introduction}
\label{sec:intro}

The discovery of strong ferromagnetism of $\rm{Ni_2MnIn}$ precipitates embedded in an antiferromagnetic (AFM) $\rm{NiMn}$ matrix~\cite{cakir_shell-ferromagnetism_2016} revealed that Heusler-based materials, which have potential applications in the areas of magnetic shape memory, magnetocaloric, and giant magnetoresistance, might possess an extremely large coercivity at room temperature. Indeed, in~\cite{scheibel_room-temperature_2017} it was demonstrated that the coercive field of such precipitates exceeds $5 \, {\rm T}$, which is remarkable for a rare-earth-free ferromagnetic (FM) material.

Detailed structural and magnetic characterization of the samples, including annealing-time and annealing-temperature studies of the segregation process~\cite{dincklage_annealing-time_2018} and FM resonance measurements~\cite{scheibel_shell-ferromagnetism_2017}, supported earlier presumptions of the existence of $5$ to $50 \, {\rm nm}$-sized inclusions that possess FM-like properties. The most significant experimentally observed feature was a vertical shift of the extracted hysteresis loop of ferromagnetic precipitates ~\cite{scheibel_room-temperature_2017} suggesting a strong exchange coupling of the precipitates to the AFM matrix. Additionaly, the shape of the extracted magnetization loop, especially its abrupt magnetization jump near zero field, suggested that the system is composed of at least two different magnetic phases. We emphasize that the described functional properties are not the unique prerogative of this compound:~similar FM properties are observed in other Heusler alloys too and are reported for $\rm{Ni_{50} Mn_{50-x}Sb_x}$ in~\cite{wanjiku_shell-ferromagnetism_2019} and for $\rm{Ni_{50} Mn_{50-x}Sn_x}$ in~\cite{cakir_transport_2020}.

The aim of this computational research is the development of a comprehensive mesoscopic micromagnetic model for materials consisting of FM and AFM phases, such as Heusler-type alloys. Mesoscopic calculations are needed due to two reasons:~first, the sizes of the FM crystallites are relatively large (in these materials up to $50 \, {\rm nm}$), rendering simulations in frames of an atomistic model not feasible due to their enormous computational effort; second, investigations of collective phenomena in such systems are highly desirable. At the same time, initial simulations at the atomic level are nevertheless necessary, because experimentally measured mesoscopic material parameters for both AFM and FM phases are lacking. Our model, which integrates both atomistic and mesoscopic simulations, enables a micromagnetic analysis of the system and furnishes a detailed quantitative account of the corresponding remagnetization processes. The crucial aspect of our study is the comparison of results obtained using our model with the previously cited experimental data. 

This two-parts article is organized as follows:~the present (first) part explains the atomistic and mesoscopic approaches to the micromagnetic model of Heusler alloys and provides all the necessary prerequisites for the three-dimensional (3D) mesoscopic calculations; the second part~\cite{Erokhin_PRB_2023_p2} contains the results of this full 3D model, and a quantitative comparison to the experimentally measured hysteresis loop of the material under study. As a first step, we conducted a thorough literature search of lattice constants, total magnetic moments, anisotropy constants, and Curie temperatures of the constituent materials provided by experimental studies and density functional theory (DFT) calculations (Sec.~\ref{sec:paraM_search}). Next, we mapped those parameters on a simplified atomic lattice structure and choose the model to describe  the exchange interaction between the different magnetic phases (Sec.~\ref{sec:struct_mapping}). In order to have the possibility to validate our results and present them in the most convenient form of magnetization profiles, we restrict ourselves at this stage to simulations of quasi~1D structures (Sec.~\ref{sec:quasi1Datomresults}). This methodology allows us to determine all the required mesoscopic parameters by comparing the magnetization distributions obtained by mesoscopic and atomistic calculations (Sec.~\ref{sec:towardmeso}). The developed approach is then generalized to 3D systems, comprising FM precipitates in an AFM matrix. 

In the second part~\cite{Erokhin_PRB_2023_p2}, we employ mesoscopic micromagnetic modeling to simulate a single FM inclusion in an AFM matrix, allowing us to reveal the details of the magnetization distribution inside both FM and AFM constituents of this system. We explicitly demonstrate that the model where the AFM matrix is treated as being \textit{monocrystalline}, is qualitatively incorrect, because it does not lead to a \textit{hysteretic} magnetization reversal process, in strong contradiction to experimental observations. To resolve this issue, we introduce a model where the AFM matrix is \textit{polycrystalline}, taking into account magnetic interactions between its different crystallites, thus obtaining the AFM/FM \textit{polycrystalline} nanocomposite. Finally, we provide a quantitative comparison between the experimentally observed magnetization loop and our simulation results, demonstrating the validity of our model.

\section{Structural and magnetic parameters}
\label{sec:paraM_search}

To find out which parameters are known reliably enough to be used in atomistic simulations, we have performed a thorough search in the literature devoted to experimental results and DFT calculations of FM and AFM phases. Experimental data obtained by x-ray diffraction demonstrate that ${\rm Ni_2MnIn}$ has a crystal structure of $L2_1$ type (cubic austenitic state) with a lattice parameter $a$ that is slightly above $6 \, \si{\angstrom}$~\cite{krenke_ferromagnetism_2006, kanomata_magnetic_2009}. A detailed study of the structural and magnetic properties~\cite{miyamoto_phase_2010}, including cumulative experimental results from various scientific groups, allows to conclude that this FM material has a low Curie temperature of $T_{\mathrm{C}} = 310 \, \rm{K}$ and a total magnetic moment of $\mu_{\rm tot} = 4.1 \, \mu_{\mathrm{B}}/{\rm f.u.}$ ($\mu_{\mathrm{B}} = $~Bohr magneton); corresponding parameters (required for the modeling) are listed in Table~\ref{table:FMexperiment}. For structural and magnetic parameters of other Heusler alloys with compositions different from that studied here, ${\rm Ni_{50}Mn_{34}In_{16}}$ and ${\rm Ni_{50}Mn_{35}In_{15}}$, but demonstrating similar magnetic properties, see \cite{umetsu_anomaly_2009} and \cite{umetsu_determination_2011}, respectively.

\begin{table}[hb]
\centering
\begin{tabular}{c|c|c|c}
 $a \, (\si{\angstrom})$ & $\mu_{\rm tot} \, (\mu_{\mathrm{B}}/{\rm f.u.})$ & $T_{\mathrm{C}} \, {\rm (K)}$ & Ref. \\ \hline
  6.071 & 4.1 & 290 & \cite{krenke_ferromagnetism_2006} \\
  6.072 & 4.1 & 300 & \cite{kanomata_magnetic_2009} \\
  6.07 & 4.2 & 310 & \cite{miyamoto_phase_2010,miyamoto_influence_2011} \\
\end{tabular}
\caption{Lattice parameter $a$, total magnetic moment $\mu_{\rm tot}$, and Curie temperature $T_{\mathrm{C}}$ of of ${\rm Ni_2MnIn}$ determined experimentally with corresponding references.}
\label{table:FMexperiment}
\end{table}

First principles calculations for this material (see Table~\ref{table:FMfirstprinciples}) provide the value of the lattice parameter $a$, which is close to the experimental data. Unfortunately,  the calculated total magnetic moment substantially depends on the particular DFT methodology. Therefore, we have relied on experimental data for $\mu_{\rm tot}$ in our simulations of this FM material.

\begin{table}[h]
\centering
\begin{tabular}{c|c|c}
 $a  \, (\si{\angstrom})$ & $\mu_{\rm tot}  \, (\mu_{\mathrm{B}}/{\rm f.u.})$ & Ref. \\ \hline
$-$ & 3.7 & \cite{godlevsky_soft_2001} \\
6.0624 & 4.22 & \cite{zayak_anomalous_2005} \\
6.053 & 4.33 & \cite{bai_first-principles_2012} \\
6.072 & 4.13 & \cite{li_role_2012} \\
$-$ & 4.3 & \cite{tan_origin_2012} \\
\end{tabular}
\caption{Lattice parameter $a$ and total magnetic moment $\mu_{\rm tot}$ of ${\rm Ni_2MnIn}$ obtained by first principles calculations.}
\label{table:FMfirstprinciples}
\end{table}

The tetragonal structure of AFM ${\rm NiMn}$ has been determined by x-ray and neutron diffraction experiments~\cite{kasper_antiferromagnetic_1959, kren_structures_1968}. In these studies it was also found that the magnetic moment of ${\rm Ni}$ atoms in this structure is almost zero, while the atomic magnetic moment on the ${\rm Mn}$ sublattices is about $4 \, \mu_{\mathrm{B}}$. Magnetic measurements~\cite{kren_structures_1968} revealed a N\'{e}el temperature of \(T_{\mathrm{N}} = 1070 \, {\rm K}\), which is extremely high compared to typical antiferromagnets ($525 \, {\rm K}$ for NiO and $293 \, {\rm K}$ for CoO). Experimentally obtained lattice parameters, magnetic moments, and N\'{e}el temperatures for ${\rm NiMn}$ are collected in Table~\ref{table:AFMexperiments}.
\begin{table}[hb]
\centering
\begin{tabular}{c|c|c|c|c}
 $a \, (\si{\angstrom})$ & $c \, (\si{\angstrom})$ & $\mu^{(+,-)}_{\rm Mn} \, (\mu_{\mathrm{B}})$ & $T_{\mathrm{N}} \, {\rm (K)}$ & Ref. \\ \hline
  3.174 & 3.524 & 4.0 & 1140 & \cite{kasper_antiferromagnetic_1959} \\
  3.74 & 3.52 & $3.8 \pm 0.3$ & $1072 \pm 40$ & \cite{kren_structures_1968} \\
\end{tabular}
\caption{Experimental parameters of ${\rm NiMn}$: lattice parameters $a$ and $c$, atomic magnetic moment $\mu$ of Mn sublattices, and N\'{e}el temperature $T_{\mathrm{N}}$.}
\label{table:AFMexperiments}
\end{table}
We also note that this AFM material is the subject of recent investigations, including the characterization of exchange-bias systems based on NiMn films~\cite{groudeva-zotova_magnetic_2003, vaskovskiy_crystal_2019} and NiMn/CoFe multilayers used in microwave applications~\cite{lamy_nimn_2006}.

First principles calculations ~\cite{sakuma_electronic_1998, schulthess_first-principles_1998, spisak_electronic_1999, nakamura_first_2003} (see Table \ref{table:AFMfirstprinciples}) yielded values for the magnetic moment of Mn atoms which are considerably smaller compared to the corresponding experimental results; hence, we have used the latter values for our atomistic simulations (see Table~\ref{table:AFMexperiments}). 

\begin{table}[h]
\centering
\begin{tabular}{c|c|c}
  $\mu^{(+,-)}_{\rm Mn} \, (\mu_{\mathrm{B}})$ & $T_{\mathrm{N}} \, {\rm (K)}$ & Ref. \\ \hline
    3.29 & $-$ & \cite{sakuma_electronic_1998} \\
    3.2 & 1187 & \cite{schulthess_first-principles_1998} \\
   3.2 (LSDA), 3.4 (GGCs) & $-$ & \cite{spisak_electronic_1999} \\
\end{tabular}
\caption{Magnetic moment $\mu$ of Mn sublattices and N\'{e}el temperature $T_{\mathrm{N}}$ of ${\rm NiMn}$ obtained by first principles calculations (LSDA:~local spin-density approximation; GGCs:~generalized gradient corrections).}
\label{table:AFMfirstprinciples}
\end{table}

We are not aware of any experimental data for anisotropy constants of the AFM phase (only the anisotropy type is known). Taking into account that these constants are essential for atomistic simulations, we have searched for {\it ab initio} calculations of corresponding values. Unfortunately, this approach faces significant challenges for {\it ab initio} theories based on the local spin-density formalism. In accordance with experiment, {\it ab initio} studies have found that the preferred orientation of magnetic moments is perpendicular to the tetragonal axis of the elementary cell, but the value of even the first-order anisotropy coefficient strongly differs in dependence on the specific {\it ab initio} method: $-1.7 \times 10^6 \, {\rm erg/cm^3}$ in~\cite{spisak_electronic_1999} versus $-9.65 \times 10^6 \, {\rm erg/cm^3}$ in~\cite{sakuma_electronic_1998}. Determination of the next-order (in-plane) magnetic anisotropy energy~\cite{spisak_electronic_1999} predicted an orientation of magnetic moments along the edges of the tetragonal cell, resulting in the in-plane anisotropy value to be only $8 \, \%$ of its out-of-plane counterpart. It was pointed out by the authors that the latter value is at the limit of accuracy of such a calculation. We have decided to use the values obtained in~\cite{spisak_electronic_1999}, as a more advanced method utilizing generalized gradient corrections is used in this study. Experimental and theoretical investigations of anisotropy in the FM phase are absent. However, a large magnetization drop at zero field seen in the hysteresis of the FM phase~\cite{scheibel_room-temperature_2017} suggests a very low anisotropy coefficient value for this phase. 

First principles calculations were also used to study the exchange stiffness in monocrystalline NiMn~\cite{nakamura_first_2003} and the corresponding intergrain exchange coupling for thin films~\cite{kai_first_2001}. The latter study demonstrated that this coupling remains significant even in the presence of relatively large spatial shifts of neighboring atomic planes.


\section{Atomic lattice structure and mapping of magnetic parameters}
\label{sec:struct_mapping}

\begin{figure}[tb!]
\centering
\resizebox{1.0\columnwidth}{!}{\includegraphics{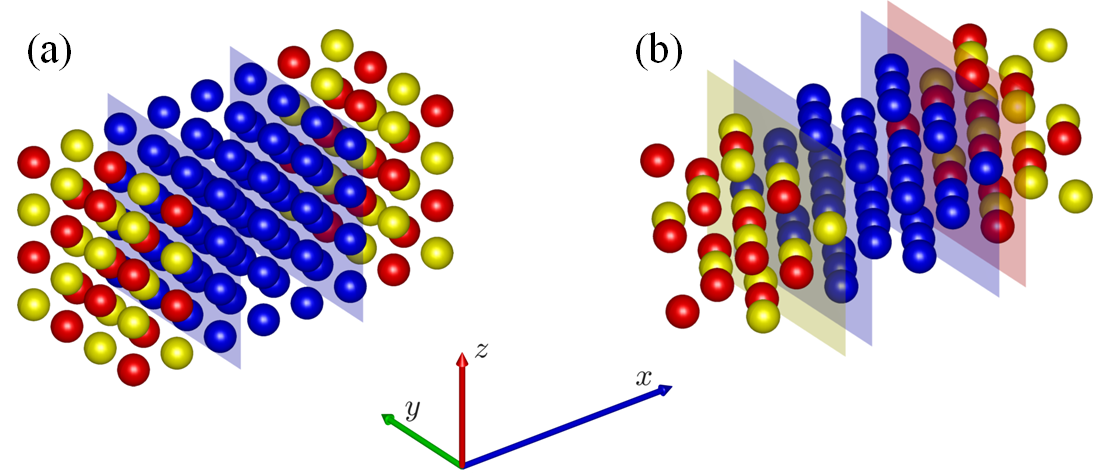}}
\caption{Part of the simulated atomistic structures containing the AFM/FM interfaces:~red and yellow spheres represent atoms belonging to different AFM sublattices; blue---atoms of the soft ferromagnet (FM inclusions). Colored planes show the planes occupied by atoms of the same type. (a)~Structure with an averaged zero interphase exchange between FM and AFM phases due to the equal fractions of atoms belonging to different sublattices of the AFM phase on the interface boundary. (b)~Structure with the maximal interphase exchange coupling on both sides of a FM inclusion. The quasi~1D structure used in actual simulations consists of $\sim$$20000$ atoms arranged on a simple cubic lattice.}
\label{figStruct1D}
\end{figure}

Simulations of a real AFM/FM interface are not feasible, because there are too many unknown structural parameters related to the corresponding interfacial disorder. For this reason we have mapped the atomistic parameters described in the previous section onto a simple cubic lattice. Examples of generated structures are presented in Fig.~\ref{figStruct1D}, where the colored spheres indicate the positions of atoms belonging to the different phases or sublattices. Red and yellow spheres represent atoms belonging to the different sublattices of the AFM, blue spheres---atoms of the soft FM phase. The aim of this figure is to illustrate that even for a simple cubic lattice there exist several possibilities to arrange the atoms at the AFM/FM interface, which will change the corresponding exchange interaction.

Note that in our model we do not explicitly introduce a pinned intermediate layer between the FM precipitates and the AFM matrix as done in \cite{cakir_shell-ferromagnetism_2016} to explain the vertical shift of the hysteresis loop. The reasons are twofold: first, the parameters of such a layer would be completely unknown, increasing the number of adjustable parameters of our model; second, such a layer is not necessary to explain neither the high coercivity of the system nor the vertical shift of the magnetization loop, as it will be shown in the next sections. Therefore, the atomic magnetic moments of the FM phase are directly coupled with the moments of the AFM phase via exchange-coupling coefficients between the atoms of different phases. 


The following atomistic results are obtained using a system of atoms arranged on a cubic lattice with dimensions of $N_x \times N_y \times N_z = 333 \times 8 \times 8$. Its length $L = 100 \, {\rm nm}$ is chosen in such a way that it can easily incorporate a domain wall (e.g., Bloch and N\'{e}el types) appearing in both AFM and FM phases. Periodic boundary conditions in all directions are applied. At the current stage, the magnetodipolar interaction is not included, because we expect that the major contribution to the system energy comes from the exchange interaction (both within the FM and AFM phases, and from the interphase exchange coupling) and the strong intrinsic anisotropy of the AFM matrix. If necessary, the  magnetodipolar interaction can added and evaluated using the lattice-Ewald method.


\begin{figure}[tb!]
\centering
\resizebox{0.85\columnwidth}{!}{\includegraphics{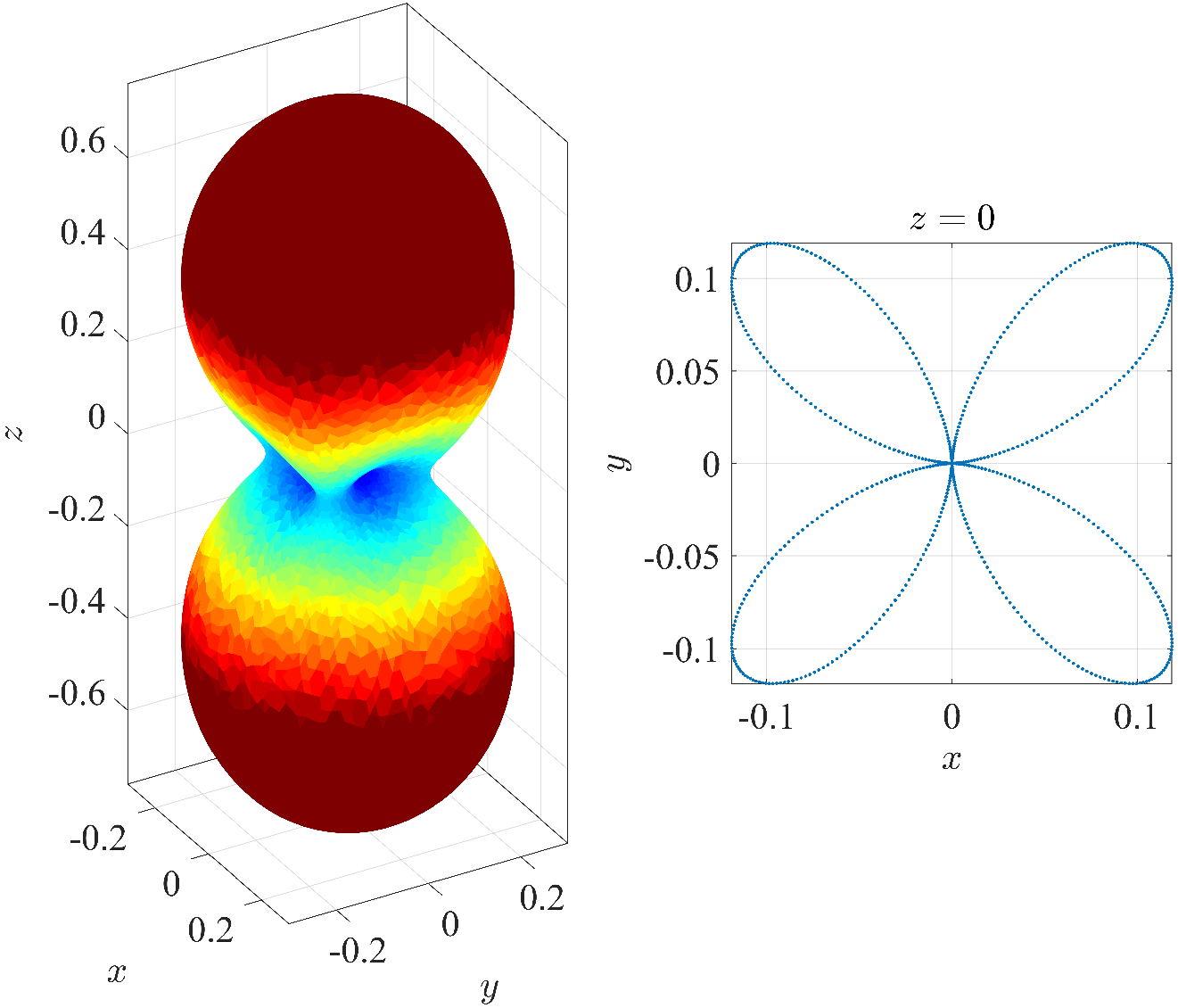}}
\caption{Tetragonal anisotropy surface (left) and its fourfold symmetry in the $x$-$y$-plane (right) obtained for parameters of a NiMn AFM.}
\label{figAniSurface}
\end{figure}


Atomistic magnetic parameters (magnetic moment $\mu_{\mathrm{s}}$ and anisotropy coefficient $k$) found in the literature for both phases (see Sec.~\ref{sec:paraM_search}) were mapped to the corresponding mesoscopic values using the data presented in Sec.~\ref{sec:paraM_search} and the standard relations
\begin{equation}
\label{eq:mom_anis_map}
\mu_{\mathrm{s}} = \frac{M_{\mathrm{s}} V_{\mathrm{a}}}{n_{\rm at}} = \frac{M_{\mathrm{s}} a^3}{n_{\rm at}}, \,\, 
k = \frac{K V_{\mathrm{a}}}{n_{\rm at}} = \frac{K a^3}{n_{\rm at}} .
\end{equation}
Here, $n_{\rm at}$ is the number of atoms per unit cell and $V_{\mathrm{a}}$ denotes the unit cell volume. For our case of a simple cubic lattice, we have $n_{\rm at} = 1$ and $V_{\mathrm{a}} = a^3$. An elementary cell size of $a = 0.3 \, {\rm nm}$ has been chosen, close to the corresponding parameter of materials under consideration. 

The anisotropy energy density of the tetragonal magnetocrystalline anisotropy of the AFM phase has the following form (see Fig.~\ref{figAniSurface}):
\begin{equation}
\label{eq:ani_energy}
\epsilon_{\mathrm{a}} = K_1 \sin^2 \theta + K_2\sin^4 \theta + K_2^{'} \sin^4 \theta \cos 4\phi, 
\end{equation}
where $\theta$ and $\phi$ are the polar and azimuthal angles. For an easy-plane anisotropy type, $K_1 < 0$ and $K_2 = 0$. The constant $K_2^{'}$ defines the basal-plane anisotropy of the fourth order. As it can be seen from comparison of the constants $K_1$ and $K_2^{'}$ (see Table~\ref{table:input_magnparam}) and from Fig.~\ref{figAniSurface}, for our AFM material the anisotropy energy is dominated by the easy-plane anisotropy and slightly disturbed by the fourfold symmetry in the $x$-$y$-plane.  For the FM phase, the standard cubic anisotropy type was chosen. As the exact value of the anisotropy constant $K_1$ of the FM phase is unknown, we have chosen a small value for this parameter, typical for a soft FM material. Table~\ref{table:input_magnparam} contains the values of magnetization and anisotropy constants used as input in our mesoscopic simulations and critical temperatures used to calculate the interatomic exchange coefficients for atomistic modeling.

\begin{table}[h]
\centering
\begin{tabular}{c|c|c}
  & $\rm{Ni_2MnIn}$ \, $\rm{(FM)}$ & $\rm{NiMn}$ \, $\rm{(AFM)}$ \\ \hline 
$M_{\mathrm{s}}$ (G) & 720 & $\pm$ 716 \\ 
anisotropy symmetry & cubic & tetragonal \\ 
$K$ (erg/cm$^3$) & $K_1 = 1.0 \times 10^4$  & $K_1 = -1.7 \times 10^6$ \\
& & $K_2^{'} = 0.136 \times 10^6$ \\ 
$T_{\rm C(N)}$ (K) & $310$ & $1070$
\end{tabular}
\caption{Mesoscopic magnetic parameters used in the simulations.}
\label{table:input_magnparam}
\end{table}

%
To determine the interatomic exchange constants $J_{ij}$, we have used the mean-field expression 
\begin{equation}
\label{eq:ex_map}
J_{ij} = \frac{3 k_{\mathrm{B}} T_{\mathrm{C}}}{z} ,
\end{equation}
which establishes the connection between these constants and the Curie temperature $T_{\mathrm{C}}$ of the ferromagnetic material; here, $k_{\mathrm{B}}$ is the Boltzmann constant and $z$ is the number of nearest neighbors in the atomic lattice (we omit the correction factor in this expression whose value is usually close to unity). For our choice of the elementary cell geometry (simple cubic lattice), $z=6$ for both FM and AFM phases. The exchange constant for the AFM material obeys the same expression, but with the N\'{e}el temperature $T_{\mathrm{N}}$ instead of $T_{\mathrm{C}}$ in Eq.~(\ref{eq:ex_map}) and the minus sign before the whole expression used due to the AFM exchange between sublattices.

Finally, the interatomic exchange coupling $J_{\rm AFM^ \leftrightarrow FM}$ between FM and AFM phases is determined in our model by the equation
%
\begin{equation}
\label{eq:FM-AFM_coupling} 
J_{\rm AFM \leftrightarrow FM} = \kappa \, \frac{J_{\rm FM}+|J_{\rm AFM}|}{2} ,
\end{equation}
%
%
where $J_{\rm FM}$ and $J_{\rm AFM}$ are the exchange constants of the FM and AFM phases and the dimensionless coefficient $0 \le \kappa \le 1$ accounts for the possible weakening of the interphase exchange due to interphase boundary imperfections. Note that for the quasi~1D atomistic simulations shown below we have used the geometry shown in Fig.~\ref{figStruct1D}, where only one of two AFM sublattices is exchange coupled to the FM phase.

\begin{figure}[tb!]
\centering
\resizebox{0.90\columnwidth}{!}{\includegraphics{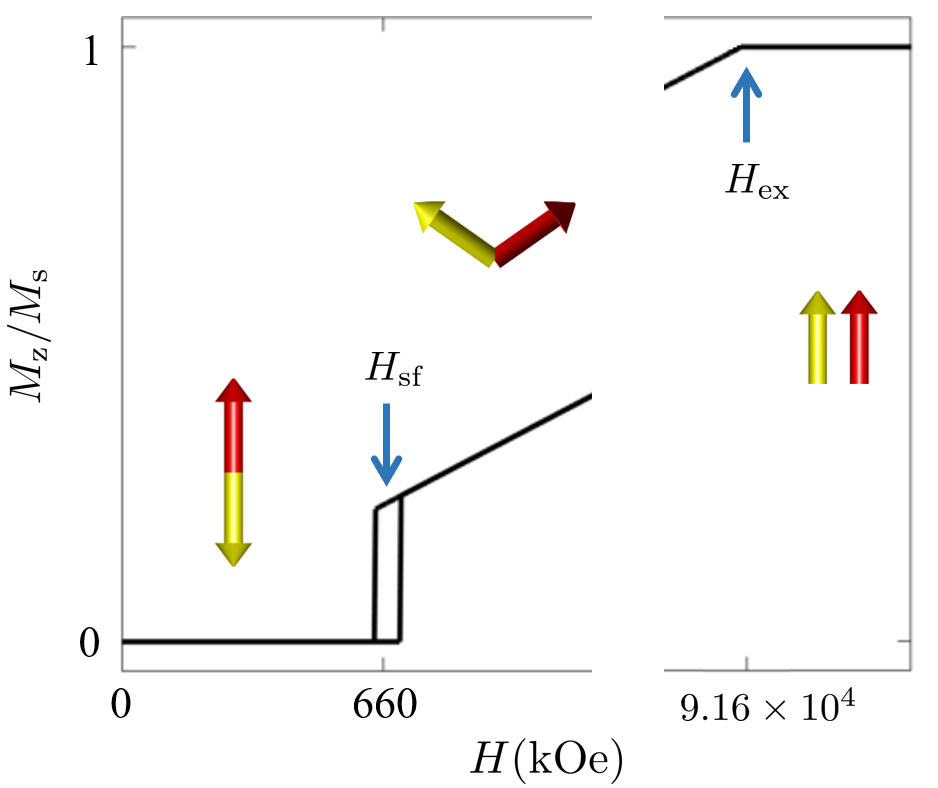}}
\caption{Magnetization curve of an uniaxial AFM with parameters taken from Table~\ref{table:input_magnparam}. Black solid line---result of the numerical modeling using our atomistic approach. Blue arrows indicate critical field values obtained analytically~\cite{vonsovskii1974book}:~spin-flop field $H_{\rm sf}$ and AFM disruption field $H_{\rm ex}$.}
\label{figAFMmagncurve}
\end{figure}

As a test for our atomistic simulations, we have compared numerically computed magnetization curves for a homogeneous uniaxial AFM with the corresponding critical fields derived analytically (see \cite{vonsovskii1974book} for details). For the parameters of the uniaxial AFM listed in Table~\ref{table:input_magnparam}, this analytical theory predicts an AFM disruption field of $H_{\rm ex}= 660 \, {\rm kOe}$ and a spin-flop field of $H_{\rm sf} = 9.2 \times 10^4 \, {\rm kOe}$. Figure~\ref{figAFMmagncurve} shows that both values are perfectly reproduced in our simulations. We note that these values are important for the further modeling: they show that if the external field does not exceed $660 \, {\rm kOe}$ ($66 \, {\rm T}$), the nearest-neighbor magnetic moments in the cubic AFM remain in the perfect antiparallel configuration. All the following results have been obtained in external fields satisfying this condition.

\begin{figure}[t]
\centering
\resizebox{0.60\columnwidth}{!}{\includegraphics{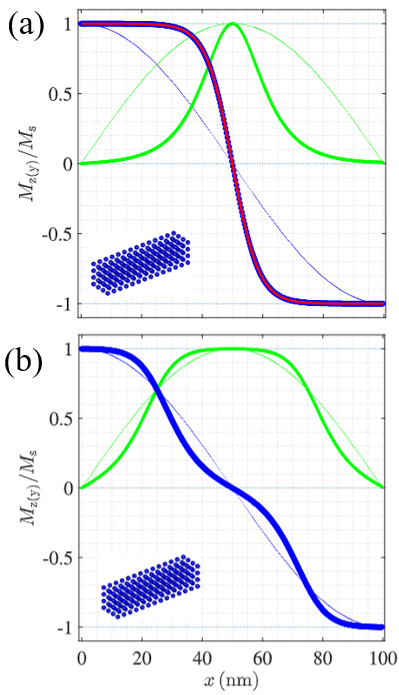}}
\caption{Magnetization profiles of Bloch walls obtained in atomistic simulations of the FM phase for the cases of (a)~uniaxial and (b)~cubic anisotropy. Solid blue lines represent the $m_z$~component of the unit magnetization vector field $\mathbf{m}$, solid green lines encode the $m_y$~component. Red line is the analytical solution for the Bloch wall in an uniaxial material (see main text). Thin blue and green lines are initial magnetization profiles (before applying the energy minimization procedure). Materials parameters for Co were assumed.}
\label{figBlochWallFM}
\end{figure}

\begin{figure}[t]
\centering
\resizebox{0.60\columnwidth}{!}{\includegraphics{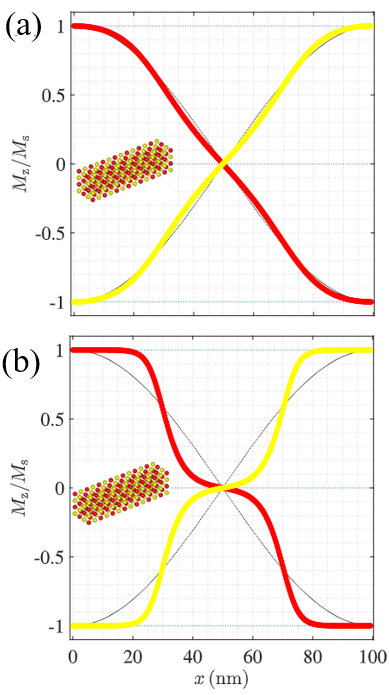}}
\caption{(a)~Magnetization profiles of Bloch walls obtained in atomistic simulations using two AFM sublattices (red and yellow lines) in NiMn with magnetic parameters taken from Table~\ref{table:input_magnparam}. (b)~The same as in (a), but with the magnetic anisotropy coefficients increased by one order of magnitude. Thin lines are initial magnetization profiles.}
\label{figBlochWallAFM}
\end{figure}


In order to validate our method for FM materials, a standard calculation of the Bloch-wall profile was carried out using the well known material parameters of Co. First, following the scheme explained above, the saturation magnetization $M_{\mathrm{s}} = 1440 \, {\rm G}$, the first-order anisotropy coefficient $K_1 = 4.1 \times 10^6 \, {\rm erg/cm^3}$, and the exchange coefficient calculated using $T_{\mathrm{C}} = 1360 \, {\rm K}$ were mapped onto a simple cubic lattice. In the starting magnetization configuration, atomic magnetic moments were arranged according to the  magnetization profile shown in Fig.~\ref{figBlochWallFM} with the thin blue line, while the moments at both ends were fixed in opposite directions. The energy minimization procedure in the absence of an external field produces a magnetization profile along the $x$~direction of the sample, which can be compared to the known analytical solution for a {\it uniaxial} anisotropy. If the anisotropy axis is in the $z$~direction, the corresponding expression has the form $M_z/M_{\mathrm{s}} = - \tanh{[\sqrt{K_1/A}(x-L/2)}]$, where $A$ is the mesoscopic exchange-stiffness constant. Plotting this expression with the standard exchange stiffness for Co, $A = 3.1 \times 10^{-6} \, {\rm erg/cm}$, we obtain a remarkable agreement between the analytical result and the atomistic simulations, as shown in Fig.~\ref{figBlochWallFM}(a). For illustrative purposes and for further comparison with the corresponding wall profile in NiMn, the same parameters were used to obtain the Bloch wall in a FM with a {\it cubic} anisotropy [Fig.~\ref{figBlochWallFM}(b)].

The same procedure as described above has been applied to the AFM NiMn with a tetragonal magnetocrystalline anisotropy and the $y$-$z$-plane chosen as the easy anisotropy plane. Figure~\ref{figBlochWallAFM} displays the magnetization profiles of the two AFM sublattices, which is qualitatively the same as for the cubic FM shown in Fig.~\ref{figBlochWallFM}(b). For a relatively small in-plane anisotropy constant, we obtain here a tanh-like Bloch-wall magnetization profile that is too close to the initial approximation (sin-like in this instance). For this reason, we had to perform additional calculations with an enlarged magnetic anisotropy constant with the result shown in Fig.~\ref{figBlochWallAFM}(b). This enlargement ensures the accuracy in the determination of the exchange-stiffness constant.

\section{Quasi one-dimensional model: Atomistic simulation results}
\label{sec:quasi1Datomresults}

At this point, all the necessary parameters for the atomistic micromagnetic simulations in  frames of our model are defined. In the following, we present selected results obtained using the quasi~1D model described above and discuss hysteresis loops and magnetization distributions at particular external fields. 

\begin{figure}[tb!]
\centering
\resizebox{0.90\columnwidth}{!}{\includegraphics{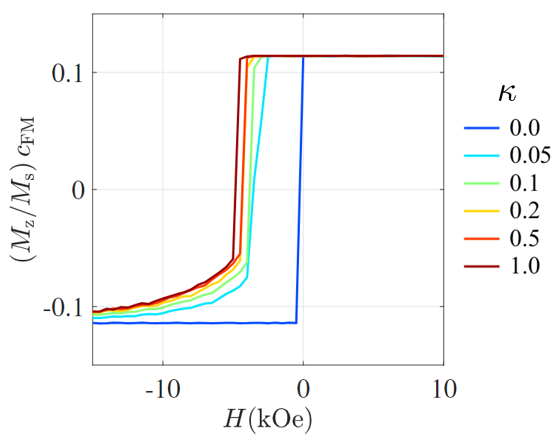}}
\caption{Remagnetization curves of the FM phase for various exchange couplings $\kappa$ between the FM inclusion (thickness:~$d_{\rm FM} = 12.6 \, {\rm nm}$) and the AFM matrix with an $y$-$z$~easy plane anisotropy.}
\label{figDepFMSize12.6nmKappaB}
\end{figure}

\begin{figure}[tb!]
\centering
\resizebox{0.90\columnwidth}{!}{\includegraphics{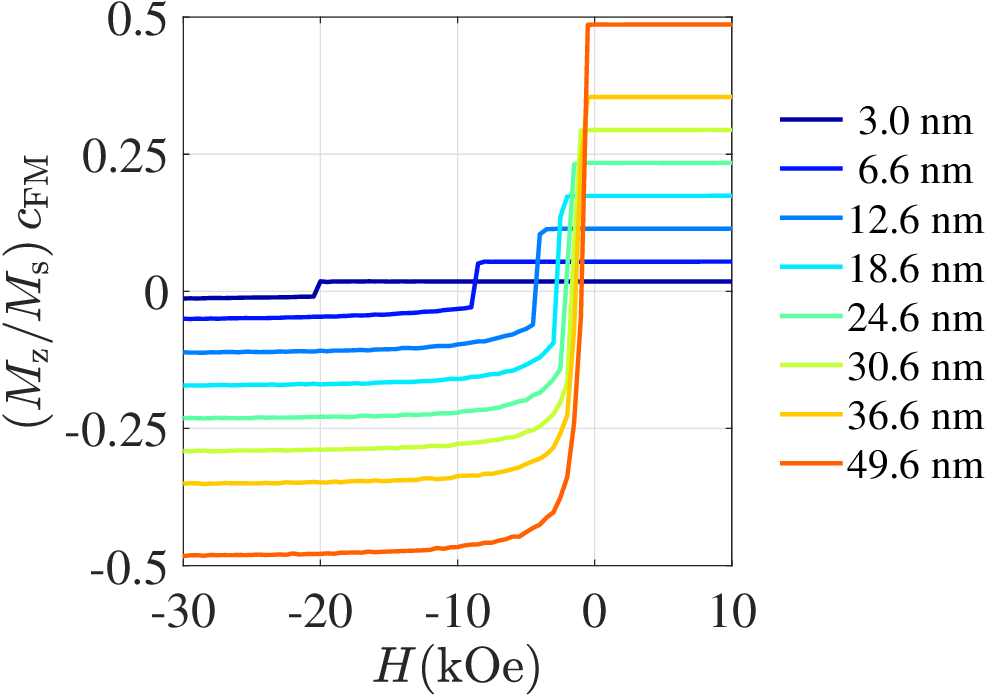}}
\caption{Remagnetization curves of the FM phase for various sizes $d_{\rm FM}$ of a FM inclusion in an AFM matrix with an $y$-$z$~easy plane anisotropy. Exchange weakening between the phases is $\kappa = 0.2$.}
\label{figDepFMSizeKappa0.2B}
\end{figure}

In our study we have varied three parameters of the model, whose variations may lead to strong changes in the magnetization reversal process: the size $d_{\rm FM}$ of the FM inclusion, the exchange weakening $\kappa$ on the AFM/FM boundary, and the orientation of the AFM anisotropy axes with respect to the external field direction, along which the $z$-axis of our coordinate system was directed. Therefore, the parameter space of our simulations is as follows (with $c_{\rm FM}$ the FM volume fraction):
\begin{itemize}
  \item size of the FM inclusion: $d_{\rm FM} = 3 \, \ldots 50 \, {\rm nm}$ \\ ($c_{\rm FM} = 3 \, \ldots 50 \, \%$)
  \item exchange weakening: $\kappa = 0.0 \, \ldots \, 1.0$
  \item anisotropy-planes orientation of the AFM phase: easy-plane anisotropies in the $x$-$z$, $y$-$z$, or $x$-$y$~coordinate planes .
\end{itemize}
Figures~\ref{figDepFMSize12.6nmKappaB} and \ref{figDepFMSizeKappa0.2B} display the results of the atomistic modeling. Hysteresis loops for different combinations of the parameters listed above are shown. The change of the hysteresis loops under the variation of the exchange weakening $\kappa$ between the AFM and FM phases is presented in Fig.~\ref{figDepFMSize12.6nmKappaB} for the example of a FM inclusion with a size of $d_{\rm FM} = 12.6 \, {\rm nm}$. Without the coupling between the phases ($\kappa = 0$), the FM inclusion reverses its magnetization at a very small negative external field due to the weak magnetocrystalline anisotropy of the FM phase. When the coupling strength increases, the coercive field increases up to $5 \, {\rm kOe}$ for this size, while for much smaller inclusions the coercivity achieves $20 \, {\rm kOe}$ (data not shown). If the coupling strength is constant (Fig.~\ref{figDepFMSizeKappa0.2B}), the coercivity decreases for larger inclusion sizes. Independent on the inclusion size, the magnetization reversal follows the same scenario:~an abrupt magnetization rotation of a significant fraction of magnetic moments of the FM phase, followed by the gradual remagnetization of remaining moments at higher fields.

The orientation of the anisotropy axes in the AFM phase with respect to the external field (see Fig.~\ref{figStruct1D}) defines the type of domain wall that is responsible for the magnetization reversal in the FM phase. For instance, a $x$-$z$~easy plane anisotropy in the AFM results in N\'{e}el-type walls in the FM inclusion, while a $y$-$z$~easy plane leads to Bloch-type walls. The third case of a $x$-$y$~easy plane demonstrates significant deviations from the perfectly aligned state already at positive fields, because in this geometry the magnetic moments are initially magnetized in the direction that is perpendicular to the easy plane (the magnetization reversal curve in this case is symmetric with a zero coercivity). The considerations presented above are valid for coupled AFM/FM systems (i.e., with $\kappa > 0$).

\begin{figure}[tb!]
\centering
\resizebox{0.90\columnwidth}{!}{\includegraphics{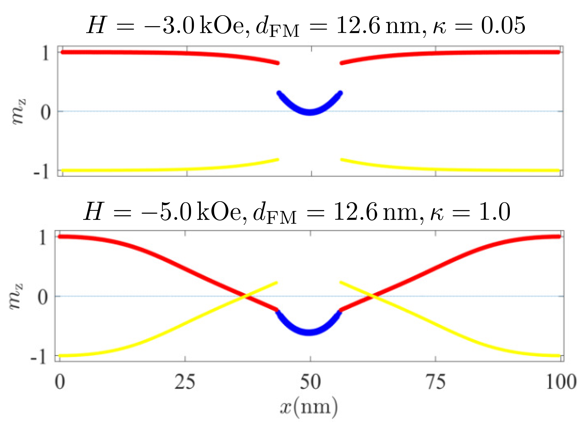}}
\caption{Magnetization profiles for the quasi~1D model at the coercive field for a structure with $d_{\rm FM}= 12.6 \, {\rm nm}$. Red and yellow lines represent the $m_z$~profiles of different AFM sublattices; blue line---$m_z$ of the soft FM inclusion. In the upper image, the exchange weakening is $\kappa = 0.05$, in the lower image $\kappa = 1.0$ (perfect coupling).}
\label{fignm12}
\end{figure}

\begin{figure}[tb!]
\centering
\resizebox{0.90\columnwidth}{!}{\includegraphics{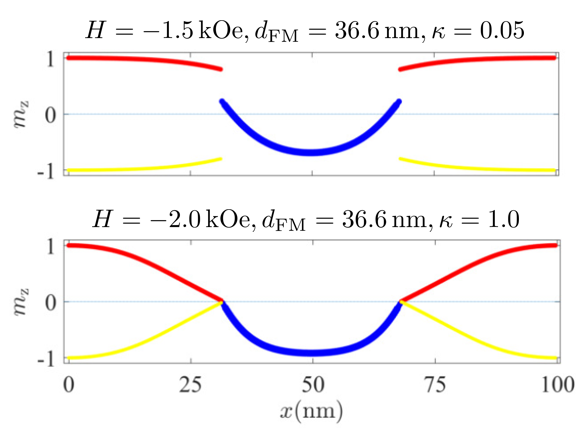}}
\caption{The same as in Fig.~\ref{fignm12}, but for a structure with $d_{\rm FM}= 36.6 \, {\rm nm}$.}
\label{fignm36}
\end{figure}

\begin{figure}[tb!]
\centering
\resizebox{1.0\columnwidth}{!}{\includegraphics{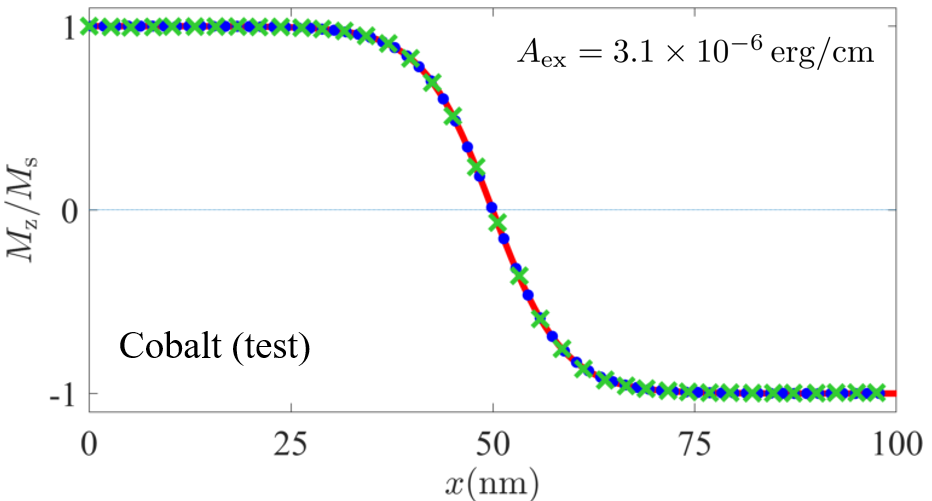}}
\caption{Test fit using atomistic and mesoscopic parameters of Co. Red line---analytical solution for uniaxial anisotropy; blue circles---atomistic simulation result; green crosses---result of the mesoscopic simulation using the exchange-stiffness constant $A_{\rm ex}$ shown in the inset, which was obtained by fitting the mesoscopic profile to the atomistic one.}
\label{figFittingAtomMesoCoTest}
\end{figure}

\begin{figure}[tb!]
\centering
\resizebox{1.0\columnwidth}{!}{\includegraphics{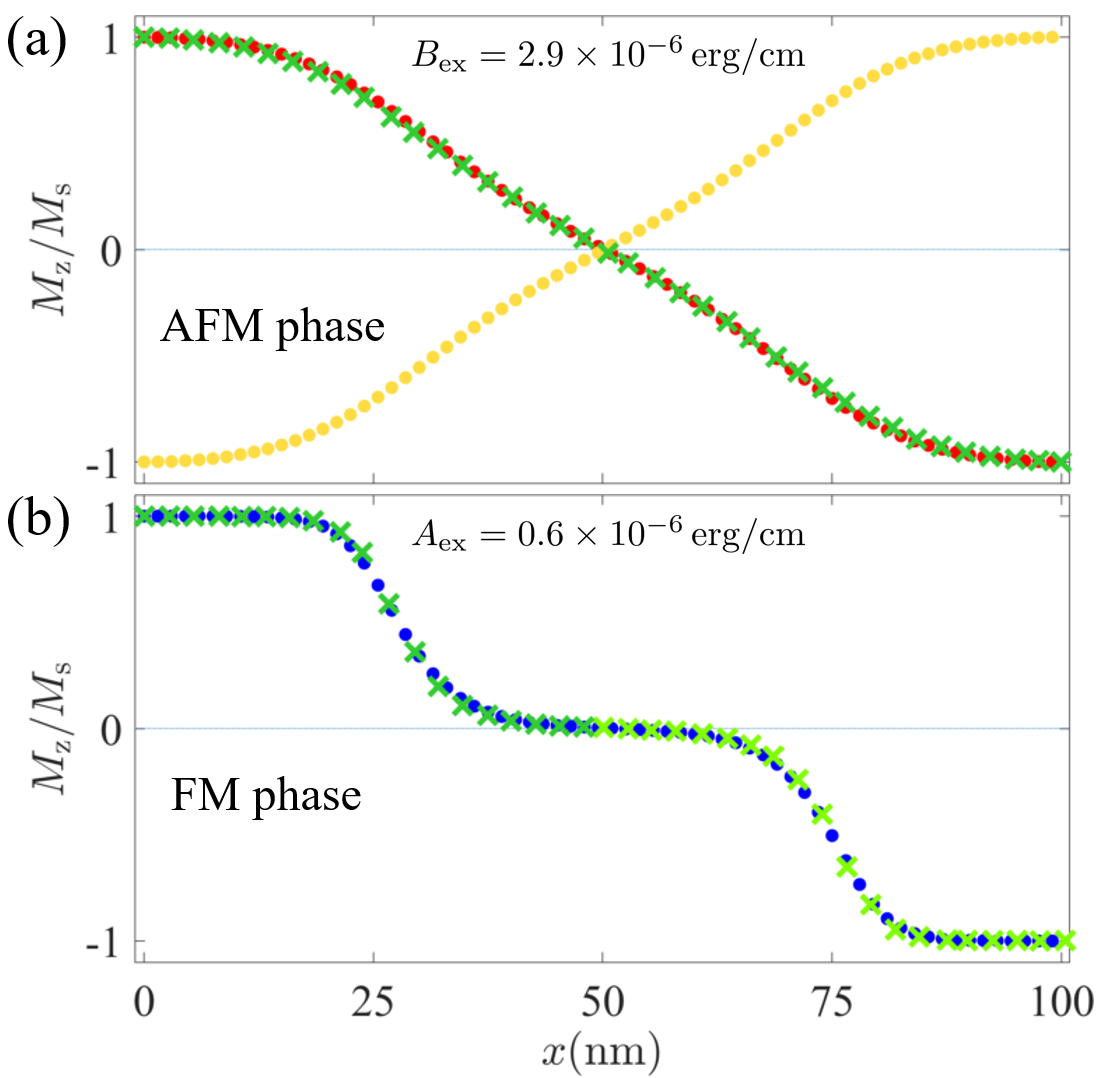}}
\caption{Results of the fitting of domain-wall profiles for the determination of the mesoscopic exchange coupling. (a)~AFM material: red and yellow circles---atomistic simulation results for Bloch walls in two AFM sublattices; green crosses---mesoscopic fit using the exchange stiffness $B$ shown in the inset. (b)~FM phase: blue circles---atomistic result; green crosses---mesoscopic fit using different spatial regions, indicating the pinning of the domain wall at the center.}
\label{figFittingAtomMesoFM_AFM}
\end{figure}

In the following, we discuss in Figs.~\ref{fignm12} and \ref{fignm36} the details of the remagnetization processes for small and large FM inclusions. Figure~\ref{fignm12} shows the magnetization distribution at the coercive field of the system with $d_{\rm FM}= 12.6 \, {\rm nm}$. The FM inclusion is almost in a single-domain state and already a strongly reduced exchange coupling ($\kappa = 0.05$) is sufficient to deflect the magnetic moments on both sides of the AFM/FM boundary. However, the coupling strength is not large enough as to eliminate the discontinuity in the magnetization distribution between the two phases. On the contrary, the magnetization profile in a system with the perfect coupling ($\kappa = 1.0$) is continuous (lower panel in Fig.~\ref{fignm12}).

An inclusion with $d_{\rm FM}= 36.6 \, {\rm nm}$, being almost three times larger than $d_{\rm FM}= 12.6 \, {\rm nm}$, can  easily incorporate a significantly inhomogeneous magnetization configuration in spite of the weak coupling (Fig.~\ref{fignm36}). The strongly coupled system ($\kappa = 1.0$) demonstrates how the magnetization state in the FM phase might significantly influence the distribution of magnetic moments in the AFM state. From the experimental point of view, such a magnetization profile is expected to give rise to a strong small-angle neutron scattering (SANS) signal~\cite{michelsbook}. Therefore, using SANS, it should be possible to obtain some (indirect) information on the magnetization state of magnetic moments belonging to the AFM phase in the neighborhood of the FM inclusion.

\section{Towards the mesoscopic approach}
\label{sec:towardmeso}

\begin{figure*}[tb!]
\centering
\resizebox{1.40\columnwidth}{!}{\includegraphics{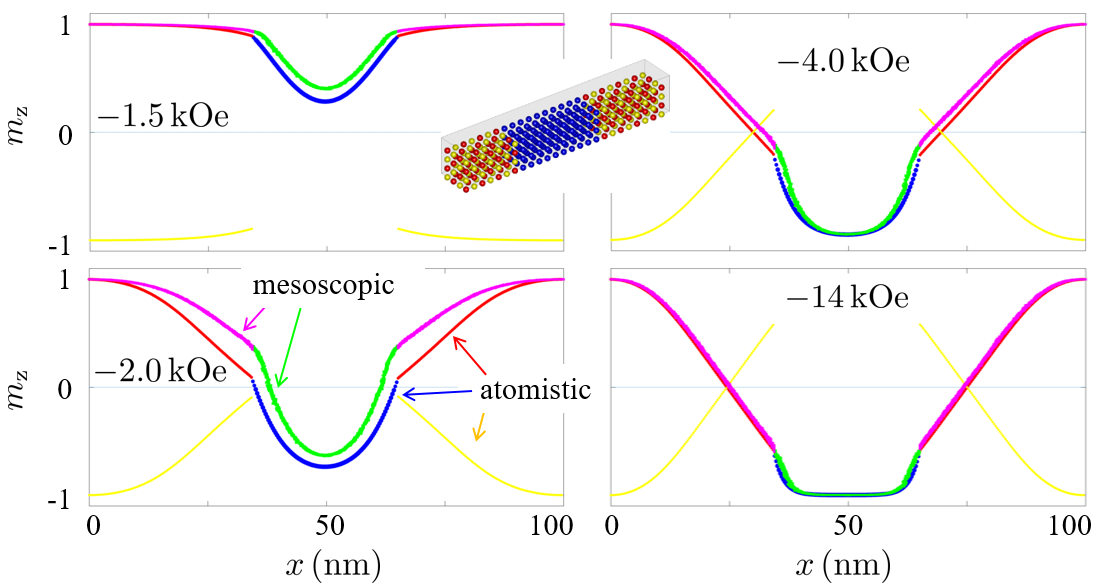}}
\caption{Comparison of atomistic and mesoscopic magnetization profiles at different external field values (see insets) ($\kappa = 1$; $d_{\rm FM} = 30 \, \rm{nm}$). Atomistic approach:~red and yellow lines correspond to the AFM sublattices; blue lines---FM phase. Mesoscopic approach: magenta line---AFM; green line---FM. Note that there is no analogue of the yellow line in the mesoscopic approach.}
\label{figProfilesAtomMeso}
\end{figure*}

\begin{figure}[tb!]
\centering
\resizebox{1.0\columnwidth}{!}{\includegraphics{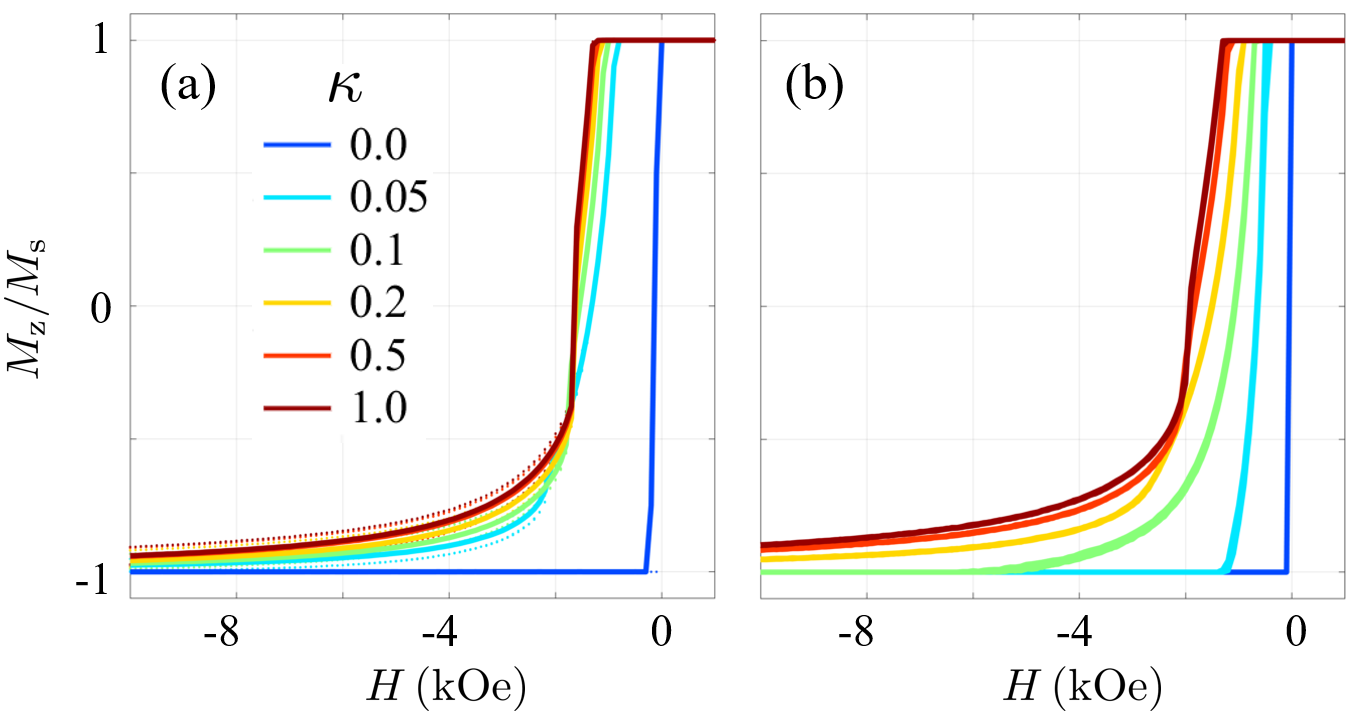}}
\caption{Comparison of the remagnetization curves of the FM phase for (a)~atomistic and (b)~mesoscopic simulations and for different values of the exchange weakening $\kappa$ on the AFM/FM interface (see legend in the left panel). Note that while the curves for large couplings ($\kappa \ge 0.5$) are quite similar, the behavior for small (but nonzero) $\kappa$ is very different.}
\label{figHystKappaAtomMeso}
\end{figure}

The typical sizes of simulated structures in atomistic simulations (especially in 3D) are limited to several hundreds of interatomic distances in each spatial direction. Therefore, for the modeling of the remagnetization process in 3D systems with sizes of hundreds of nanometers, the transition to the mesoscopic length scale is mandatory. In this section, we describe the steps necessary to introduce and validate the corresponding mesoscopic model.

First, in order to implement a mesoscopic model for an AFM, it is necessary to introduce an effective vector that describes the direction of the magnetic moments in the AFM sublattices with a spatial resolution of a few nanometers (analogous to the standard procedure in micromagnetics). This vector coincides with the magnetization direction of one of the sublattices and is antiparallel to the moments of the second sublattice (under the above described limitation of the external field magnitude). In the following discussion, we will refer to this vector as a {\it descriptor} of the spatial distribution of magnetization in the AFM phase.

Second, in analogy to conventional micromagnetics, we have to define the size of finite elements having a constant ``magnetization'' (usually a few nanometers, $3.5 \, \rm{nm}$ in our case) and all the interaction coefficients both within one phase and between the different phases. The mesoscopic saturation magnetization and anisotropy coefficient for a FM and for both sublattices of an AFM can be derived from their atomistic counterparts using Eqs.~(\ref{eq:mom_anis_map}). In order to obtain the mesoscopic exchange-stiffness constant $A_{\rm ex}$ and its analogue for the AFM phase $B_{\rm ex}$, it is necessary to employ a special fitting procedure due to the (weakly) nonlocal nature of the exchange interaction. This procedure consists of finding such an exchange-stiffness constant for which the mesoscopically simulated Bloch-wall profile coincides with its atomistic counterpart. 

As a test calculation (see Fig.~\ref{figFittingAtomMesoCoTest}), we have used the magnetic materials parameters of Co, obtaining again for the exchange-stiffness constant the textbook value of $3.1 \times 10^{-6} \, {\rm erg/cm}$. Figure~\ref{figFittingAtomMesoFM_AFM} presents the results of the fittings for AFM and FM phases, giving the values $B_{\rm ex} = 2.9 \times 10^{-6} \, {\rm erg/cm}$ for the AFM phase and $A_{\rm ex} = 0.6 \times 10^{-6} \, {\rm erg/cm}$ for FM phase. In the latter case an enlarged anisotropy constant was used in order to increase the accuracy of the fit (see the discussion related to Fig.~\ref{figBlochWallAFM}). We have also pinned the domain wall at the center of the simulated region ($x = 50 \, \rm{nm}$) to keep the magnetization profile symmetric.

Next, we have compared the magnetization profiles obtained by the atomistic and mesoscopic modeling for the system containing a FM inclusion within the AFM phase. This comparison is of special importance, because it allows us to draw the conclusion about the reliability of the mesoscopic approach for the following simulations. Corresponding plots are shown in  Fig.~\ref{figProfilesAtomMeso}. We note that atomistic and mesoscopic calculations were performed independently, only using the magnetic materials parameters corresponding to the respective spatial scale. As a result, we achieved an excellent match between atomistic and mesoscopic approaches that allows us to study the magnetization distribution in AFM/FM composites at a length scale of the order of hundreds of nanometers, while resolving its details at the nanometer scale.

Special attention deserves the relation between the exchange-weakening coefficients $\kappa$ present in both the atomistic and mesoscopic methods. This relation is highly nontrivial, because these coefficients lead in different approaches to the weakening of the exchange coupling between two very different ``building blocks'' of corresponding models. We have revealed this relation by comparing remagnetization curves obtained by both methods. The result is presented in Fig.~\ref{figHystKappaAtomMeso}, where the same values for the exchange-weakening coefficient $0 \le \kappa \le 1$ for the two approaches have been used.


\section{Conclusion}

In the first of two articles, we have presented a combined atomistic/mesoscopic approach for micromagnetic simulations of systems containing coupled ferromagnetic (FM) and antiferromagnetic (AFM) phases. The reliability of the atomistic simulations was tested using single-phase FM and AFM materials, where we have simulated Bloch-wall magnetization profiles and compared them with analytical results for the same system. Further, the developed methodology was validated by comparing the results of mesoscopic simulations with the corresponding ones of atomistic modeling for quasi one-dimensional systems, being either single-phase or consisting of a FM inclusion within an AFM phase. In the second part~\cite{Erokhin_PRB_2023_p2}, we will extend our simulation methodology to polycrystalline systems and show that the experimentally found large coercivity of $\sim$$5 \, \mathrm{T}$ for a Heusler-type alloy~\cite{scheibel_room-temperature_2017} can be naturally explained by the pinning of the FM phase via the extremely hard AFM. 


\bibliography{AFM_Heusler}

\begin{thebibliography}{30}%
\makeatletter
\providecommand \@ifxundefined [1]{%
 \@ifx{#1\undefined}
}%
\providecommand \@ifnum [1]{%
 \ifnum #1\expandafter \@firstoftwo
 \else \expandafter \@secondoftwo
 \fi
}%
\providecommand \@ifx [1]{%
 \ifx #1\expandafter \@firstoftwo
 \else \expandafter \@secondoftwo
 \fi
}%
\providecommand \natexlab [1]{#1}%
\providecommand \enquote  [1]{``#1''}%
\providecommand \bibnamefont  [1]{#1}%
\providecommand \bibfnamefont [1]{#1}%
\providecommand \citenamefont [1]{#1}%
\providecommand \href@noop [0]{\@secondoftwo}%
\providecommand \href [0]{\begingroup \@sanitize@url \@href}%
\providecommand \@href[1]{\@@startlink{#1}\@@href}%
\providecommand \@@href[1]{\endgroup#1\@@endlink}%
\providecommand \@sanitize@url [0]{\catcode `\\12\catcode `\$12\catcode
  `\&12\catcode `\#12\catcode `\^12\catcode `\_12\catcode `\%12\relax}%
\providecommand \@@startlink[1]{}%
\providecommand \@@endlink[0]{}%
\providecommand \url  [0]{\begingroup\@sanitize@url \@url }%
\providecommand \@url [1]{\endgroup\@href {#1}{\urlprefix }}%
\providecommand \urlprefix  [0]{URL }%
\providecommand \Eprint [0]{\href }%
\providecommand \doibase [0]{https://doi.org/}%
\providecommand \selectlanguage [0]{\@gobble}%
\providecommand \bibinfo  [0]{\@secondoftwo}%
\providecommand \bibfield  [0]{\@secondoftwo}%
\providecommand \translation [1]{[#1]}%
\providecommand \BibitemOpen [0]{}%
\providecommand \bibitemStop [0]{}%
\providecommand \bibitemNoStop [0]{.\EOS\space}%
\providecommand \EOS [0]{\spacefactor3000\relax}%
\providecommand \BibitemShut  [1]{\csname bibitem#1\endcsname}%
\let\auto@bib@innerbib\@empty
\bibitem [{\citenamefont {Çakır}\ \emph {et~al.}(2016)\citenamefont
  {Çakır}, \citenamefont {Acet},\ and\ \citenamefont
  {Farle}}]{cakir_shell-ferromagnetism_2016}%
  \BibitemOpen
  \bibfield  {author} {\bibinfo {author} {\bibfnamefont {A.}~\bibnamefont
  {Çakır}}, \bibinfo {author} {\bibfnamefont {M.}~\bibnamefont {Acet}},\ and\
  \bibinfo {author} {\bibfnamefont {M.}~\bibnamefont {Farle}},\ }\href
  {http://www.nature.czom/articles/srep28931} {\bibfield  {journal} {\bibinfo
  {journal} {Sci. Rep.}\ }\textbf {\bibinfo {volume} {6}},\ \bibinfo {pages}
  {28931} (\bibinfo {year} {2016})}\BibitemShut {NoStop}%
\bibitem [{\citenamefont {Scheibel}\ \emph
  {et~al.}(2017{\natexlab{a}})\citenamefont {Scheibel}, \citenamefont
  {Spoddig}, \citenamefont {Meckenstock}, \citenamefont {Gottschall},
  \citenamefont {Çakır}, \citenamefont {Krenke}, \citenamefont {Farle},
  \citenamefont {Gutfleisch},\ and\ \citenamefont
  {Acet}}]{scheibel_room-temperature_2017}%
  \BibitemOpen
  \bibfield  {author} {\bibinfo {author} {\bibfnamefont {F.}~\bibnamefont
  {Scheibel}}, \bibinfo {author} {\bibfnamefont {D.}~\bibnamefont {Spoddig}},
  \bibinfo {author} {\bibfnamefont {R.}~\bibnamefont {Meckenstock}}, \bibinfo
  {author} {\bibfnamefont {T.}~\bibnamefont {Gottschall}}, \bibinfo {author}
  {\bibfnamefont {A.}~\bibnamefont {Çakır}}, \bibinfo {author} {\bibfnamefont
  {T.}~\bibnamefont {Krenke}}, \bibinfo {author} {\bibfnamefont
  {M.}~\bibnamefont {Farle}}, \bibinfo {author} {\bibfnamefont
  {O.}~\bibnamefont {Gutfleisch}},\ and\ \bibinfo {author} {\bibfnamefont
  {M.}~\bibnamefont {Acet}},\ }\href {https://doi.org/10.1063/1.4983199}
  {\bibfield  {journal} {\bibinfo  {journal} {Appl. Phys. Lett.}\ }\textbf
  {\bibinfo {volume} {110}},\ \bibinfo {pages} {192406} (\bibinfo {year}
  {2017}{\natexlab{a}})}\BibitemShut {NoStop}%
\bibitem [{\citenamefont {Dincklage}\ \emph {et~al.}(2018)\citenamefont
  {Dincklage}, \citenamefont {Scheibel}, \citenamefont {Çakır}, \citenamefont
  {Farle},\ and\ \citenamefont {Acet}}]{dincklage_annealing-time_2018}%
  \BibitemOpen
  \bibfield  {author} {\bibinfo {author} {\bibfnamefont {L.}~\bibnamefont
  {Dincklage}}, \bibinfo {author} {\bibfnamefont {F.}~\bibnamefont {Scheibel}},
  \bibinfo {author} {\bibfnamefont {A.}~\bibnamefont {Çakır}}, \bibinfo
  {author} {\bibfnamefont {M.}~\bibnamefont {Farle}},\ and\ \bibinfo {author}
  {\bibfnamefont {M.}~\bibnamefont {Acet}},\ }\href
  {https://doi.org/10.1063/1.5018851} {\bibfield  {journal} {\bibinfo
  {journal} {{AIP} Advances}\ }\textbf {\bibinfo {volume} {8}},\ \bibinfo
  {pages} {025012} (\bibinfo {year} {2018})}\BibitemShut {NoStop}%
\bibitem [{\citenamefont {Scheibel}\ \emph
  {et~al.}(2017{\natexlab{b}})\citenamefont {Scheibel}, \citenamefont
  {Spoddig}, \citenamefont {Meckenstock}, \citenamefont {Çakır},
  \citenamefont {Farle},\ and\ \citenamefont
  {Acet}}]{scheibel_shell-ferromagnetism_2017}%
  \BibitemOpen
  \bibfield  {author} {\bibinfo {author} {\bibfnamefont {F.}~\bibnamefont
  {Scheibel}}, \bibinfo {author} {\bibfnamefont {D.}~\bibnamefont {Spoddig}},
  \bibinfo {author} {\bibfnamefont {R.}~\bibnamefont {Meckenstock}}, \bibinfo
  {author} {\bibfnamefont {A.}~\bibnamefont {Çakır}}, \bibinfo {author}
  {\bibfnamefont {M.}~\bibnamefont {Farle}},\ and\ \bibinfo {author}
  {\bibfnamefont {M.}~\bibnamefont {Acet}},\ }\href
  {https://doi.org/10.1063/1.4976335} {\bibfield  {journal} {\bibinfo
  {journal} {{AIP} Advances}\ }\textbf {\bibinfo {volume} {7}},\ \bibinfo
  {pages} {056425} (\bibinfo {year} {2017}{\natexlab{b}})}\BibitemShut
  {NoStop}%
\bibitem [{\citenamefont {Wanjiku}\ \emph {et~al.}(2019)\citenamefont
  {Wanjiku}, \citenamefont {Çakır}, \citenamefont {Scheibel}, \citenamefont
  {Wiedwald}, \citenamefont {Farle},\ and\ \citenamefont
  {Acet}}]{wanjiku_shell-ferromagnetism_2019}%
  \BibitemOpen
  \bibfield  {author} {\bibinfo {author} {\bibfnamefont {Z.}~\bibnamefont
  {Wanjiku}}, \bibinfo {author} {\bibfnamefont {A.}~\bibnamefont {Çakır}},
  \bibinfo {author} {\bibfnamefont {F.}~\bibnamefont {Scheibel}}, \bibinfo
  {author} {\bibfnamefont {U.}~\bibnamefont {Wiedwald}}, \bibinfo {author}
  {\bibfnamefont {M.}~\bibnamefont {Farle}},\ and\ \bibinfo {author}
  {\bibfnamefont {M.}~\bibnamefont {Acet}},\ }\href
  {https://doi.org/10.1063/1.5057763} {\bibfield  {journal} {\bibinfo
  {journal} {Journal of Applied Physics}\ }\textbf {\bibinfo {volume} {125}},\
  \bibinfo {pages} {043902} (\bibinfo {year} {2019})}\BibitemShut {NoStop}%
\bibitem [{\citenamefont {Çakır}\ \emph {et~al.}(2020)\citenamefont
  {Çakır}, \citenamefont {Koyun}, \citenamefont {Acet},\ and\ \citenamefont
  {Farle}}]{cakir_transport_2020}%
  \BibitemOpen
  \bibfield  {author} {\bibinfo {author} {\bibfnamefont {A.}~\bibnamefont
  {Çakır}}, \bibinfo {author} {\bibfnamefont {H.~N.}\ \bibnamefont {Koyun}},
  \bibinfo {author} {\bibfnamefont {M.}~\bibnamefont {Acet}},\ and\ \bibinfo
  {author} {\bibfnamefont {M.}~\bibnamefont {Farle}},\ }\href
  {https://doi.org/10.1016/j.jmmm.2019.166265} {\bibfield  {journal} {\bibinfo
  {journal} {Journal of Magnetism and Magnetic Materials}\ }\textbf {\bibinfo
  {volume} {499}},\ \bibinfo {pages} {166265} (\bibinfo {year}
  {2020})}\BibitemShut {NoStop}%
\bibitem [{\citenamefont {Erokhin}\ \emph {et~al.}(2023)\citenamefont
  {Erokhin}, \citenamefont {Berkov},\ and\ \citenamefont
  {Michels}}]{Erokhin_PRB_2023_p2}%
  \BibitemOpen
  \bibfield  {author} {\bibinfo {author} {\bibfnamefont {S.}~\bibnamefont
  {Erokhin}}, \bibinfo {author} {\bibfnamefont {D.}~\bibnamefont {Berkov}},\
  and\ \bibinfo {author} {\bibfnamefont {A.}~\bibnamefont {Michels}},\ }\href
  {https://doi.org/10.1103/PhysRevB.85.024410} {\bibfield  {journal} {\bibinfo
  {journal} {Phys. Rev. B}\ }\textbf {\bibinfo {volume} {108}},\ \bibinfo
  {pages} {xyzabc} (\bibinfo {year} {2023})}\BibitemShut {NoStop}%
\bibitem [{\citenamefont {Krenke}\ \emph {et~al.}(2006)\citenamefont {Krenke},
  \citenamefont {Acet}, \citenamefont {Wassermann}, \citenamefont {Moya},
  \citenamefont {Mañosa},\ and\ \citenamefont
  {Planes}}]{krenke_ferromagnetism_2006}%
  \BibitemOpen
  \bibfield  {author} {\bibinfo {author} {\bibfnamefont {T.}~\bibnamefont
  {Krenke}}, \bibinfo {author} {\bibfnamefont {M.}~\bibnamefont {Acet}},
  \bibinfo {author} {\bibfnamefont {E.~F.}\ \bibnamefont {Wassermann}},
  \bibinfo {author} {\bibfnamefont {X.}~\bibnamefont {Moya}}, \bibinfo {author}
  {\bibfnamefont {L.}~\bibnamefont {Mañosa}},\ and\ \bibinfo {author}
  {\bibfnamefont {A.}~\bibnamefont {Planes}},\ }\bibfield  {journal} {\bibinfo
  {journal} {Physical Review B}\ }\textbf {\bibinfo {volume} {73}},\ \href
  {https://doi.org/10.1103/PhysRevB.73.174413} {10.1103/PhysRevB.73.174413}
  (\bibinfo {year} {2006})\BibitemShut {NoStop}%
\bibitem [{\citenamefont {Kanomata}\ \emph {et~al.}(2009)\citenamefont
  {Kanomata}, \citenamefont {Yasuda}, \citenamefont {Sasaki}, \citenamefont
  {Nishihara}, \citenamefont {Kainuma}, \citenamefont {Ito}, \citenamefont
  {Oikawa}, \citenamefont {Ishida}, \citenamefont {Neumann},\ and\
  \citenamefont {Ziebeck}}]{kanomata_magnetic_2009}%
  \BibitemOpen
  \bibfield  {author} {\bibinfo {author} {\bibfnamefont {T.}~\bibnamefont
  {Kanomata}}, \bibinfo {author} {\bibfnamefont {T.}~\bibnamefont {Yasuda}},
  \bibinfo {author} {\bibfnamefont {S.}~\bibnamefont {Sasaki}}, \bibinfo
  {author} {\bibfnamefont {H.}~\bibnamefont {Nishihara}}, \bibinfo {author}
  {\bibfnamefont {R.}~\bibnamefont {Kainuma}}, \bibinfo {author} {\bibfnamefont
  {W.}~\bibnamefont {Ito}}, \bibinfo {author} {\bibfnamefont {K.}~\bibnamefont
  {Oikawa}}, \bibinfo {author} {\bibfnamefont {K.}~\bibnamefont {Ishida}},
  \bibinfo {author} {\bibfnamefont {K.-U.}\ \bibnamefont {Neumann}},\ and\
  \bibinfo {author} {\bibfnamefont {K.}~\bibnamefont {Ziebeck}},\ }\href
  {https://doi.org/10.1016/j.jmmm.2008.11.079} {\bibfield  {journal} {\bibinfo
  {journal} {Journal of Magnetism and Magnetic Materials}\ }\textbf {\bibinfo
  {volume} {321}},\ \bibinfo {pages} {773} (\bibinfo {year}
  {2009})}\BibitemShut {NoStop}%
\bibitem [{\citenamefont {Miyamoto}\ \emph {et~al.}(2010)\citenamefont
  {Miyamoto}, \citenamefont {Ito}, \citenamefont {Umetsu}, \citenamefont
  {Kainuma}, \citenamefont {Kanomata},\ and\ \citenamefont
  {Ishida}}]{miyamoto_phase_2010}%
  \BibitemOpen
  \bibfield  {author} {\bibinfo {author} {\bibfnamefont {T.}~\bibnamefont
  {Miyamoto}}, \bibinfo {author} {\bibfnamefont {W.}~\bibnamefont {Ito}},
  \bibinfo {author} {\bibfnamefont {R.}~\bibnamefont {Umetsu}}, \bibinfo
  {author} {\bibfnamefont {R.}~\bibnamefont {Kainuma}}, \bibinfo {author}
  {\bibfnamefont {T.}~\bibnamefont {Kanomata}},\ and\ \bibinfo {author}
  {\bibfnamefont {K.}~\bibnamefont {Ishida}},\ }\href
  {https://doi.org/10.1016/j.scriptamat.2009.10.006} {\bibfield  {journal}
  {\bibinfo  {journal} {Scripta Materialia}\ }\textbf {\bibinfo {volume}
  {62}},\ \bibinfo {pages} {151} (\bibinfo {year} {2010})}\BibitemShut
  {NoStop}%
\bibitem [{\citenamefont {Umetsu}\ \emph {et~al.}(2009)\citenamefont {Umetsu},
  \citenamefont {Ito}, \citenamefont {Ito}, \citenamefont {Koyama},
  \citenamefont {Fujita}, \citenamefont {Oikawa}, \citenamefont {Kanomata},
  \citenamefont {Kainuma},\ and\ \citenamefont {Ishida}}]{umetsu_anomaly_2009}%
  \BibitemOpen
  \bibfield  {author} {\bibinfo {author} {\bibfnamefont {R.}~\bibnamefont
  {Umetsu}}, \bibinfo {author} {\bibfnamefont {W.}~\bibnamefont {Ito}},
  \bibinfo {author} {\bibfnamefont {K.}~\bibnamefont {Ito}}, \bibinfo {author}
  {\bibfnamefont {K.}~\bibnamefont {Koyama}}, \bibinfo {author} {\bibfnamefont
  {A.}~\bibnamefont {Fujita}}, \bibinfo {author} {\bibfnamefont
  {K.}~\bibnamefont {Oikawa}}, \bibinfo {author} {\bibfnamefont
  {T.}~\bibnamefont {Kanomata}}, \bibinfo {author} {\bibfnamefont
  {R.}~\bibnamefont {Kainuma}},\ and\ \bibinfo {author} {\bibfnamefont
  {K.}~\bibnamefont {Ishida}},\ }\href
  {https://doi.org/10.1016/j.scriptamat.2008.08.022} {\bibfield  {journal}
  {\bibinfo  {journal} {Scripta Materialia}\ }\textbf {\bibinfo {volume}
  {60}},\ \bibinfo {pages} {25} (\bibinfo {year} {2009})}\BibitemShut {NoStop}%
\bibitem [{\citenamefont {Umetsu}\ \emph {et~al.}(2011)\citenamefont {Umetsu},
  \citenamefont {Fujita}, \citenamefont {Ito}, \citenamefont {Kanomata},\ and\
  \citenamefont {Kainuma}}]{umetsu_determination_2011}%
  \BibitemOpen
  \bibfield  {author} {\bibinfo {author} {\bibfnamefont {R.~Y.}\ \bibnamefont
  {Umetsu}}, \bibinfo {author} {\bibfnamefont {A.}~\bibnamefont {Fujita}},
  \bibinfo {author} {\bibfnamefont {W.}~\bibnamefont {Ito}}, \bibinfo {author}
  {\bibfnamefont {T.}~\bibnamefont {Kanomata}},\ and\ \bibinfo {author}
  {\bibfnamefont {R.}~\bibnamefont {Kainuma}},\ }\href
  {https://doi.org/10.1088/0953-8984/23/32/326001} {\bibfield  {journal}
  {\bibinfo  {journal} {Journal of Physics: Condensed Matter}\ }\textbf
  {\bibinfo {volume} {23}},\ \bibinfo {pages} {326001} (\bibinfo {year}
  {2011})}\BibitemShut {NoStop}%
\bibitem [{\citenamefont {Miyamoto}\ \emph {et~al.}(2011)\citenamefont
  {Miyamoto}, \citenamefont {Ito}, \citenamefont {Umetsu}, \citenamefont
  {Kanomata}, \citenamefont {Ishida},\ and\ \citenamefont
  {Kainuma}}]{miyamoto_influence_2011}%
  \BibitemOpen
  \bibfield  {author} {\bibinfo {author} {\bibfnamefont {T.}~\bibnamefont
  {Miyamoto}}, \bibinfo {author} {\bibfnamefont {W.}~\bibnamefont {Ito}},
  \bibinfo {author} {\bibfnamefont {R.~Y.}\ \bibnamefont {Umetsu}}, \bibinfo
  {author} {\bibfnamefont {T.}~\bibnamefont {Kanomata}}, \bibinfo {author}
  {\bibfnamefont {K.}~\bibnamefont {Ishida}},\ and\ \bibinfo {author}
  {\bibfnamefont {R.}~\bibnamefont {Kainuma}},\ }\href
  {https://doi.org/10.2320/matertrans.M2011125} {\bibfield  {journal} {\bibinfo
   {journal} {{MATERIALS} {TRANSACTIONS}}\ }\textbf {\bibinfo {volume} {52}},\
  \bibinfo {pages} {1836} (\bibinfo {year} {2011})}\BibitemShut {NoStop}%
\bibitem [{\citenamefont {Godlevsky}\ and\ \citenamefont
  {Rabe}(2001)}]{godlevsky_soft_2001}%
  \BibitemOpen
  \bibfield  {author} {\bibinfo {author} {\bibfnamefont {V.~V.}\ \bibnamefont
  {Godlevsky}}\ and\ \bibinfo {author} {\bibfnamefont {K.~M.}\ \bibnamefont
  {Rabe}},\ }\bibfield  {journal} {\bibinfo  {journal} {Physical Review B}\
  }\textbf {\bibinfo {volume} {63}},\ \href
  {https://doi.org/10.1103/PhysRevB.63.134407} {10.1103/PhysRevB.63.134407}
  (\bibinfo {year} {2001})\BibitemShut {NoStop}%
\bibitem [{\citenamefont {Zayak}\ \emph {et~al.}(2005)\citenamefont {Zayak},
  \citenamefont {Entel}, \citenamefont {Rabe}, \citenamefont {Adeagbo},\ and\
  \citenamefont {Acet}}]{zayak_anomalous_2005}%
  \BibitemOpen
  \bibfield  {author} {\bibinfo {author} {\bibfnamefont {A.~T.}\ \bibnamefont
  {Zayak}}, \bibinfo {author} {\bibfnamefont {P.}~\bibnamefont {Entel}},
  \bibinfo {author} {\bibfnamefont {K.~M.}\ \bibnamefont {Rabe}}, \bibinfo
  {author} {\bibfnamefont {W.~A.}\ \bibnamefont {Adeagbo}},\ and\ \bibinfo
  {author} {\bibfnamefont {M.}~\bibnamefont {Acet}},\ }\bibfield  {journal}
  {\bibinfo  {journal} {Physical Review B}\ }\textbf {\bibinfo {volume} {72}},\
  \href {https://doi.org/10.1103/PhysRevB.72.054113}
  {10.1103/PhysRevB.72.054113} (\bibinfo {year} {2005})\BibitemShut {NoStop}%
\bibitem [{\citenamefont {Bai}\ \emph {et~al.}(2012)\citenamefont {Bai},
  \citenamefont {Xu}, \citenamefont {Raulot}, \citenamefont {Zhang},
  \citenamefont {Esling}, \citenamefont {Zhao},\ and\ \citenamefont
  {Zuo}}]{bai_first-principles_2012}%
  \BibitemOpen
  \bibfield  {author} {\bibinfo {author} {\bibfnamefont {J.}~\bibnamefont
  {Bai}}, \bibinfo {author} {\bibfnamefont {N.}~\bibnamefont {Xu}}, \bibinfo
  {author} {\bibfnamefont {J.-M.}\ \bibnamefont {Raulot}}, \bibinfo {author}
  {\bibfnamefont {Y.~D.}\ \bibnamefont {Zhang}}, \bibinfo {author}
  {\bibfnamefont {C.}~\bibnamefont {Esling}}, \bibinfo {author} {\bibfnamefont
  {X.}~\bibnamefont {Zhao}},\ and\ \bibinfo {author} {\bibfnamefont
  {L.}~\bibnamefont {Zuo}},\ }\href {https://doi.org/10.1063/1.4767331}
  {\bibfield  {journal} {\bibinfo  {journal} {Journal of Applied Physics}\
  }\textbf {\bibinfo {volume} {112}},\ \bibinfo {pages} {114901} (\bibinfo
  {year} {2012})}\BibitemShut {NoStop}%
\bibitem [{\citenamefont {Li}\ \emph {et~al.}(2012)\citenamefont {Li},
  \citenamefont {Luo}, \citenamefont {Hu}, \citenamefont {Yang}, \citenamefont
  {Johansson},\ and\ \citenamefont {Vitos}}]{li_role_2012}%
  \BibitemOpen
  \bibfield  {author} {\bibinfo {author} {\bibfnamefont {C.-M.}\ \bibnamefont
  {Li}}, \bibinfo {author} {\bibfnamefont {H.-B.}\ \bibnamefont {Luo}},
  \bibinfo {author} {\bibfnamefont {Q.-M.}\ \bibnamefont {Hu}}, \bibinfo
  {author} {\bibfnamefont {R.}~\bibnamefont {Yang}}, \bibinfo {author}
  {\bibfnamefont {B.}~\bibnamefont {Johansson}},\ and\ \bibinfo {author}
  {\bibfnamefont {L.}~\bibnamefont {Vitos}},\ }\bibfield  {journal} {\bibinfo
  {journal} {Physical Review B}\ }\textbf {\bibinfo {volume} {86}},\ \href
  {https://doi.org/10.1103/PhysRevB.86.214205} {10.1103/PhysRevB.86.214205}
  (\bibinfo {year} {2012})\BibitemShut {NoStop}%
\bibitem [{\citenamefont {Tan}\ \emph {et~al.}(2012)\citenamefont {Tan},
  \citenamefont {Huang}, \citenamefont {Tian}, \citenamefont {Jiang},\ and\
  \citenamefont {Cai}}]{tan_origin_2012}%
  \BibitemOpen
  \bibfield  {author} {\bibinfo {author} {\bibfnamefont {C.~L.}\ \bibnamefont
  {Tan}}, \bibinfo {author} {\bibfnamefont {Y.~W.}\ \bibnamefont {Huang}},
  \bibinfo {author} {\bibfnamefont {X.~H.}\ \bibnamefont {Tian}}, \bibinfo
  {author} {\bibfnamefont {J.~X.}\ \bibnamefont {Jiang}},\ and\ \bibinfo
  {author} {\bibfnamefont {W.}~\bibnamefont {Cai}},\ }\href
  {https://doi.org/10.1063/1.3697637} {\bibfield  {journal} {\bibinfo
  {journal} {Applied Physics Letters}\ }\textbf {\bibinfo {volume} {100}},\
  \bibinfo {pages} {132402} (\bibinfo {year} {2012})}\BibitemShut {NoStop}%
\bibitem [{\citenamefont {Kasper}\ and\ \citenamefont
  {Kouvel}(1959)}]{kasper_antiferromagnetic_1959}%
  \BibitemOpen
  \bibfield  {author} {\bibinfo {author} {\bibfnamefont {J.}~\bibnamefont
  {Kasper}}\ and\ \bibinfo {author} {\bibfnamefont {J.}~\bibnamefont
  {Kouvel}},\ }\href {https://doi.org/10.1016/0022-3697(59)90219-7} {\bibfield
  {journal} {\bibinfo  {journal} {Journal of Physics and Chemistry of Solids}\
  }\textbf {\bibinfo {volume} {11}},\ \bibinfo {pages} {231} (\bibinfo {year}
  {1959})}\BibitemShut {NoStop}%
\bibitem [{\citenamefont {Krén}\ \emph {et~al.}(1968)\citenamefont {Krén},
  \citenamefont {Nagy}, \citenamefont {Nagy}, \citenamefont {Pál},\ and\
  \citenamefont {Szabó}}]{kren_structures_1968}%
  \BibitemOpen
  \bibfield  {author} {\bibinfo {author} {\bibfnamefont {E.}~\bibnamefont
  {Krén}}, \bibinfo {author} {\bibfnamefont {E.}~\bibnamefont {Nagy}},
  \bibinfo {author} {\bibfnamefont {I.}~\bibnamefont {Nagy}}, \bibinfo {author}
  {\bibfnamefont {L.}~\bibnamefont {Pál}},\ and\ \bibinfo {author}
  {\bibfnamefont {P.}~\bibnamefont {Szabó}},\ }\href
  {https://doi.org/https://doi.org/10.1016/0022-3697(68)90259-X} {\bibfield
  {journal} {\bibinfo  {journal} {Journal of Physics and Chemistry of Solids}\
  }\textbf {\bibinfo {volume} {29}},\ \bibinfo {pages} {101 } (\bibinfo {year}
  {1968})}\BibitemShut {NoStop}%
\bibitem [{\citenamefont {Groudeva-Zotova}\ \emph {et~al.}(2003)\citenamefont
  {Groudeva-Zotova}, \citenamefont {Elefant}, \citenamefont {Kaltofen},
  \citenamefont {Tietjen}, \citenamefont {Thomas}, \citenamefont {Hoffmann},\
  and\ \citenamefont {Schneider}}]{groudeva-zotova_magnetic_2003}%
  \BibitemOpen
  \bibfield  {author} {\bibinfo {author} {\bibfnamefont {S.}~\bibnamefont
  {Groudeva-Zotova}}, \bibinfo {author} {\bibfnamefont {D.}~\bibnamefont
  {Elefant}}, \bibinfo {author} {\bibfnamefont {R.}~\bibnamefont {Kaltofen}},
  \bibinfo {author} {\bibfnamefont {D.}~\bibnamefont {Tietjen}}, \bibinfo
  {author} {\bibfnamefont {J.}~\bibnamefont {Thomas}}, \bibinfo {author}
  {\bibfnamefont {V.}~\bibnamefont {Hoffmann}},\ and\ \bibinfo {author}
  {\bibfnamefont {C.}~\bibnamefont {Schneider}},\ }\href
  {https://doi.org/10.1016/S0304-8853(02)01535-4} {\bibfield  {journal}
  {\bibinfo  {journal} {Journal of Magnetism and Magnetic Materials}\ }\textbf
  {\bibinfo {volume} {263}},\ \bibinfo {pages} {57} (\bibinfo {year}
  {2003})}\BibitemShut {NoStop}%
\bibitem [{\citenamefont {Vas'kovskiy}\ \emph {et~al.}(2019)\citenamefont
  {Vas'kovskiy}, \citenamefont {Moskalev}, \citenamefont {Lepalovskij},
  \citenamefont {Svalov}, \citenamefont {Larrañaga}, \citenamefont {Balymov},\
  and\ \citenamefont {Kulesh}}]{vaskovskiy_crystal_2019}%
  \BibitemOpen
  \bibfield  {author} {\bibinfo {author} {\bibfnamefont {V.}~\bibnamefont
  {Vas'kovskiy}}, \bibinfo {author} {\bibfnamefont {M.}~\bibnamefont
  {Moskalev}}, \bibinfo {author} {\bibfnamefont {V.}~\bibnamefont
  {Lepalovskij}}, \bibinfo {author} {\bibfnamefont {A.}~\bibnamefont {Svalov}},
  \bibinfo {author} {\bibfnamefont {A.}~\bibnamefont {Larrañaga}}, \bibinfo
  {author} {\bibfnamefont {K.}~\bibnamefont {Balymov}},\ and\ \bibinfo {author}
  {\bibfnamefont {N.}~\bibnamefont {Kulesh}},\ }\href
  {https://doi.org/10.1016/j.jallcom.2018.11.016} {\bibfield  {journal}
  {\bibinfo  {journal} {Journal of Alloys and Compounds}\ }\textbf {\bibinfo
  {volume} {777}},\ \bibinfo {pages} {264} (\bibinfo {year}
  {2019})}\BibitemShut {NoStop}%
\bibitem [{\citenamefont {Lamy}\ and\ \citenamefont
  {Viala}(2006)}]{lamy_nimn_2006}%
  \BibitemOpen
  \bibfield  {author} {\bibinfo {author} {\bibfnamefont {Y.}~\bibnamefont
  {Lamy}}\ and\ \bibinfo {author} {\bibfnamefont {B.}~\bibnamefont {Viala}},\
  }\href {https://doi.org/10.1109/TMAG.2006.878871} {\bibfield  {journal}
  {\bibinfo  {journal} {{IEEE} Transactions on Magnetics}\ }\textbf {\bibinfo
  {volume} {42}},\ \bibinfo {pages} {3332} (\bibinfo {year}
  {2006})}\BibitemShut {NoStop}%
\bibitem [{\citenamefont {Sakuma}(1998)}]{sakuma_electronic_1998}%
  \BibitemOpen
  \bibfield  {author} {\bibinfo {author} {\bibfnamefont {A.}~\bibnamefont
  {Sakuma}},\ }\href {https://doi.org/10.1016/S0304-8853(98)00115-2} {\bibfield
   {journal} {\bibinfo  {journal} {Journal of Magnetism and Magnetic
  Materials}\ }\textbf {\bibinfo {volume} {187}},\ \bibinfo {pages} {105}
  (\bibinfo {year} {1998})}\BibitemShut {NoStop}%
\bibitem [{\citenamefont {Schulthess}\ and\ \citenamefont
  {Butler}(1998)}]{schulthess_first-principles_1998}%
  \BibitemOpen
  \bibfield  {author} {\bibinfo {author} {\bibfnamefont {T.~C.}\ \bibnamefont
  {Schulthess}}\ and\ \bibinfo {author} {\bibfnamefont {W.~H.}\ \bibnamefont
  {Butler}},\ }\href {https://doi.org/10.1063/1.367824} {\bibfield  {journal}
  {\bibinfo  {journal} {Journal of Applied Physics}\ }\textbf {\bibinfo
  {volume} {83}},\ \bibinfo {pages} {7225} (\bibinfo {year}
  {1998})}\BibitemShut {NoStop}%
\bibitem [{\citenamefont {Spisák}\ and\ \citenamefont
  {Hafner}(1999)}]{spisak_electronic_1999}%
  \BibitemOpen
  \bibfield  {author} {\bibinfo {author} {\bibfnamefont {D.}~\bibnamefont
  {Spisák}}\ and\ \bibinfo {author} {\bibfnamefont {J.}~\bibnamefont
  {Hafner}},\ }\href {https://doi.org/10.1088/0953-8984/11/33/306} {\bibfield
  {journal} {\bibinfo  {journal} {Journal of Physics: Condensed Matter}\
  }\textbf {\bibinfo {volume} {11}},\ \bibinfo {pages} {6359} (\bibinfo {year}
  {1999})}\BibitemShut {NoStop}%
\bibitem [{\citenamefont {Nakamura}\ \emph {et~al.}(2003)\citenamefont
  {Nakamura}, \citenamefont {Ito}, \citenamefont {Freeman}, \citenamefont
  {Zhong},\ and\ \citenamefont {Fernandez-de Castro}}]{nakamura_first_2003}%
  \BibitemOpen
  \bibfield  {author} {\bibinfo {author} {\bibfnamefont {K.}~\bibnamefont
  {Nakamura}}, \bibinfo {author} {\bibfnamefont {T.}~\bibnamefont {Ito}},
  \bibinfo {author} {\bibfnamefont {A.~J.}\ \bibnamefont {Freeman}}, \bibinfo
  {author} {\bibfnamefont {L.}~\bibnamefont {Zhong}},\ and\ \bibinfo {author}
  {\bibfnamefont {J.}~\bibnamefont {Fernandez-de Castro}},\ }\href
  {https://doi.org/10.1063/1.1556152} {\bibfield  {journal} {\bibinfo
  {journal} {Journal of Applied Physics}\ }\textbf {\bibinfo {volume} {93}},\
  \bibinfo {pages} {6879} (\bibinfo {year} {2003})}\BibitemShut {NoStop}%
\bibitem [{\citenamefont {Kai}\ \emph {et~al.}(2001)\citenamefont {Kai},
  \citenamefont {Fujiwara}, \citenamefont {Schulthess},\ and\ \citenamefont
  {Butler}}]{kai_first_2001}%
  \BibitemOpen
  \bibfield  {author} {\bibinfo {author} {\bibfnamefont {T.}~\bibnamefont
  {Kai}}, \bibinfo {author} {\bibfnamefont {H.}~\bibnamefont {Fujiwara}},
  \bibinfo {author} {\bibfnamefont {T.~C.}\ \bibnamefont {Schulthess}},\ and\
  \bibinfo {author} {\bibfnamefont {W.~H.}\ \bibnamefont {Butler}},\ }\href
  {https://doi.org/10.1063/1.1371945} {\bibfield  {journal} {\bibinfo
  {journal} {Journal of Applied Physics}\ }\textbf {\bibinfo {volume} {89}},\
  \bibinfo {pages} {7940} (\bibinfo {year} {2001})}\BibitemShut {NoStop}%
\bibitem [{\citenamefont {Vonsovksii}(1974)}]{vonsovskii1974book}%
  \BibitemOpen
  \bibfield  {author} {\bibinfo {author} {\bibfnamefont {S.}~\bibnamefont
  {Vonsovksii}},\ }\href@noop {} {\emph {\bibinfo {title} {{Magnetism}}}}\
  (\bibinfo  {publisher} {Wiley},\ \bibinfo {address} {New York},\ \bibinfo
  {year} {1974})\ Chap.~\bibinfo {chapter} {22}\BibitemShut {NoStop}%
\bibitem [{\citenamefont {Michels}(2021)}]{michelsbook}%
  \BibitemOpen
  \bibfield  {author} {\bibinfo {author} {\bibfnamefont {A.}~\bibnamefont
  {Michels}},\ }\href@noop {} {\emph {\bibinfo {title} {{Magnetic Small-Angle
  Neutron Scattering: A Probe for Mesoscale Magnetism Analysis}}}}\ (\bibinfo
  {publisher} {Oxford University Press},\ \bibinfo {address} {Oxford},\
  \bibinfo {year} {2021})\BibitemShut {NoStop}%
\end{thebibliography}%

\end{document}